\documentstyle[preprint]{ptptex}
\notypesetlogo
\title{D$3$-brane from D-instantons in  the hybrid formalism
of superstring}
\author{Yoshiteru 
 {\sc Amemiya}\footnote{e-mail: amemiya@het.ph.tsukuba.ac.jp}, 
 Kenji {\sc Araki}\footnote{e-mail: araki@het.ph.tsukuba.ac.jp}
 and So {\sc Katagiri}\footnote{e-mail: katagiri@het.ph.tsukuba.ac.jp}}
\inst{Institute of Physics, University of Tsukuba, Ibaraki 305-8571, Japan}
\abst{
A D$p$-brane  can be regarded as a configuration of  infinitely many
D$(p-2k)$-branes in bosonic string. 
We will show  this property of D-branes in the superstring case using
 the hybrid formalism.
It is convenient to study   the boundary state for such D-branes 
to  study such property between D-branes.
We show that the boundary state for a  D$3$-brane in a  
constant self-dual gauge field
 background can be expressed in terms of  the boundary state for
 D-instantons in the hybrid formalism.
}
\begin{document}
\maketitle

\section {Introduction}

Field theories on the noncommutative spacetime have 
been studied extensively. In the bosonic string theory, the
noncommutativity of the bosonic coordinates arises at the end points of
the open string on a D-brane in a
constant NS-NS two form field  background~\cite{aas,ch}. 
This noncommutativity can be understood from a different point of view. 
A D$p$-brane with a constant gauge field background can be regarded as a
configuration of infinitely many D$(p-2k)$-branes~\cite{is1,is2}. 
From this point of view, the noncommutativity is 
expressed as the noncommutativity of the matrix coordinates 
of D$(p-2k)$-branes. 
It is convenient to study   the boundary state for such D-branes 
to  study such property between D-branes.
Indeed, we can see from the boundary state that the 
Seiberg-Witten map~\cite{sw} between
commutative and noncommutative  gauge fields can be regarded
as the map between  two different pictures of the D-brane~\cite{ok}.
Furthermore, this point of view supports  matrix models which treat
lower dimensional D-branes as
the fundamental degrees of freedom.

Recently, field theories on the deformed
superspace have been considered, in which fermionic coordinates are
non(anti)commutative~\cite{si,besi}. In the superstring theory, this nonanticommutativity
arises in a constant self-dual graviphoton background~\cite{si,besi,bgn,ov}. The graviphoton
field  corresponds to a RR vertex operator. Therefore, 
the hybrid formalism~\cite{be} instead of the NSR formalism 
is useful to study such a background.
The non(anti)commutativity 
 of the fermionic coordinates arise at the end points of
the open string.

We expect that  
the fermionic non(anti)commutativity can also be understood  by
using lower dimensional D-branes. If we can do so, 
we  may be able  to discuss the  
Seiberg-Witten map~\cite{mi} and supermatrix models~\cite{pe,kkm,hiu,st}. 
As a first step, we consider whether the  boundary state for a
D$p$-brane in a constant  gauge field background can be represented 
by that of  a configuration of
infinitely many D-instantons, using  the hybrid variables.   
  
This paper is organized as follows.
In section 2, we review how to  express the relation between
 a D$p$-brane and infinitely many
D$(p-2k)$-branes in the bosonic string theory. 
In section 3, we show that the boundary state for a  D$3$-brane in a  
constant self-dual gauge field
 background can be expressed in terms of  the boundary state for
 D-instantons in the hybrid superstring.  
Section 4 is devoted to conclusions and discussions.

\section {D$p$-brane from D($p-2k$)-branes }

In this section, we review how a D$p$-brane can be expressed as
a configuration of 
infinitely many D($p-2k$)-branes. In order to show this fact, the easiest
way is to study the boundary state.
For simplicity, we treat  the case of $p=1,k=1$. 
It is straightforward to generalize to other cases.

Let us consider the boundary state for a D1-brane in 
the bosonic string. A configuration of infinitely many
D-instantons is represented by $\infty \times \infty$ matrices $X^m$.
 We will consider the configuration of 
D-instantons such as
\begin{eqnarray}
X^1=\hat M^1,\\\nonumber
X^2=\hat M^2,
\end{eqnarray}
where the matrices $\hat M^1,\hat M^2 $ satisfy
\begin{eqnarray}
[\hat M^1,\hat M^2]=i\theta.
\end{eqnarray}
Here $\theta$ is a constant.
The boundary state corresponding to this configuration is 
expressed as
\begin{equation}
\label{b1}
 |B\big>=TrP\exp(-i\int_{0}^{2\pi}p^m (\sigma)\hat M_m)|B\big>_{-1},
\end{equation} 
where $P$ denotes the path ordering with respect to $\sigma$ and
$p^m$ is the conjugate momentum to $x^m$.   $|B\big>_{-1}$ denotes
the boundary state for a D-instanton 
at the origin, $x^m(\sigma)|B\big>_{-1}=0$.  
This boundary state can be rewritten as a path integral representation,
	\begin{equation}
\label{12}
|B\big>=\int[dY^1 dY^2]\exp[{i\over{\theta}}\int d\sigma Y^2\partial_\sigma
Y^1-i\int d\sigma (p^1 Y^1 +p^2 Y^2)]|B\big>_{-1}.
 \end{equation}
It is easy to see that this boundary state coincides
with the boundary state for a D1-brane with a constant gauge field 
background $F_{21}={1\over \theta}$. 
Indeed, the following
identities hold:
\begin{eqnarray}
0&=&
\int[dY^1 dY^2]{\delta\over{\delta Y^1}}\exp[{i\over{\theta}}\int d\sigma Y^2\partial_\sigma
Y^1-i\int d\sigma (p^1 Y^1 +p^2 Y^2)]|B\big>_{-1}\\\nonumber
&=&[-{i\over{\theta}}\partial_{\sigma}x^2-ip^1]|B\big>,\\\nonumber
0&=&
\int[dY^1 dY^2]{\delta\over{\delta Y^2}}\exp[{i\over{\theta}}\int d\sigma Y^1\partial_\sigma
Y^2-i\int d\sigma (p^1 Y^1 +p^2 Y^2)]|B\big>_{-1}\\\nonumber
&=&[{i\over{\theta}}\partial_{\sigma}x^1-ip^2]|B\big>.
\end{eqnarray}
Therefore, this boundary state  coincides with 
\begin{equation}
\label{b6}
\exp[{i\over\theta}\int d\sigma
(x^1\partial_{\sigma}x^2)]|B\big>_1,
\end{equation}
 up to normalization,
where $|B\big>_1$ denotes the boundary state for a D1-brane  
satisfying ~$p_1|B\big>_1=p_2|B\big>_1=x_3|B\big>_1=\cdots=x_D|B\big>_1=0$.

\section { D$3$-brane from D-instantons in the hybrid formalism}

In this section, we show that  the boundary state for a 
D$3$-brane  in a constant
self-dual gauge field background can be expressed in terms of
the  boundary state for infinitely many D-instantons in the hybrid formalism.

 Here, we use the hybrid 
superstring on $R^4\times X^6$~\cite{be}.
The world-sheet action\footnote{We use the convention of ref.[8]} in a constant
self-dual gauge field background is given by
\begin{equation}
 S={1\over\pi}\int d\tau d\sigma 
({1\over2}\partial y^m\tilde\partial y_m-q_\alpha\tilde\partial\theta^\alpha 
+\bar d_{\dot\alpha}\tilde\partial\bar\theta^{\dot\alpha}-{1\over2}\partial\rho\tilde\partial\rho
-\tilde{q}_\alpha\partial\tilde{\theta^\alpha} 
+\tilde{\bar d_{\dot\alpha}}\partial\tilde{\bar\theta^{\dot\alpha}}
-{1\over2}\partial\tilde\rho\tilde\partial\tilde\rho)+S_C+S_F,
\end{equation}
where $S_C$ is the  action for the  variables corresponding to $X^6$.
 $S_F$ represents 
 a constant self-dual gauge field background, which is given by
\begin{eqnarray}
\label{ver}
 S_F=-{i\over2}\int d\sigma F_{mn}\big\{
y^m\partial_\sigma y^n
+iq^\alpha\sigma^m_{\alpha\dot\alpha}
{\bar\sigma}^{n\dot\alpha\beta}\theta_\beta
-i\tilde{q}^\alpha\sigma^m_{\alpha\dot\alpha}
{\bar\sigma}^{n\dot\alpha\beta}\tilde{\theta}_\beta
\}.
\end{eqnarray}

Before we go on, we need to mention the equations satisfied by 
the boundary state  for a  D$3$-brane. In a flat background, 
the boundary state $|B\big>_{3f}$ satisfies the following equations,
\begin{eqnarray}
\label{d5}
(\partial+\tilde\partial)y^m|B\big>_{3f}=0,\\\nonumber
(q^\alpha+\tilde q^\alpha)|B\big>_{3f}=(\theta^\alpha-\tilde
\theta^\alpha)|B\big>_{3f}=0,\\\nonumber
(\bar d^{\dot\alpha}
+\tilde{\bar d^{\dot\alpha}})|B\big>_{3f}=(\bar\theta^{\dot\alpha}-\tilde{\bar
{\theta^{\dot\alpha}}})|B\big>_{3f}=0,\\\nonumber
(e^{\rho}+e^{\tilde\rho})|B\big>_{3f}=0.
\end{eqnarray}
In a constant self-dual gauge field background,
the equations for  the bosonic fields are modified as
\begin{equation}
\label{d3b}
\{{1\over 4\pi}(\partial+\tilde\partial)y^m+F_{mn} (\partial-\tilde\partial)y^n\}|B\big>_3=0.
\end{equation}
We can obtain the equations for  the fermionic fields 
from the equations for the NSR variables  because
the map between the hybrid variables and the NSR variables is known.
Here instead, we utilize the N=2 world-sheet supersymmetry
to  determine the  equations for the
fermionic fields.
 The N=2 supersymmetry algebra  consists of  the generators,
\begin{eqnarray}
 T&=&{1\over2}\partial y^m\partial y^m-q_\alpha\partial\theta^\alpha +
\bar{d}_{\dot\alpha}\partial\bar{\theta}^{\dot\alpha}
-{1\over2}\partial \rho\partial \rho+{1\over2}\partial^2 \rho
+T_C+{i\over2}\partial J_C,\\\nonumber
G^+&=&
e^{\rho}d_\alpha d^\alpha+G^+_C,\\\nonumber
G^-&=&e^{-\rho}\bar{d}_{\dot\alpha}\bar{d}^{\dot\alpha}
+G^-_C,\\\nonumber
J&=&i\partial\rho+J_C,
\end{eqnarray}
where
\begin{equation}
 d_\alpha=q_\alpha+2i\sigma^n_{\alpha\dot\alpha}\bar\theta^{\dot\alpha}\partial y_n
-4\bar{\theta}\bar{\theta}\partial\theta_\alpha,
\end{equation}
and $ [T_C,G^+_C,G^-_C,J_C]$ are the c=9 N=2 superconformal
generators of the compactified space $X^6$. 
We  determine the  equations for the
fermionic fields  using the fact that the boundary state should
preserve this symmetry. 
Since the boundary action in eq.(\ref{ver}) 
does not involve $\rho$, the boundary state $|B\big>_{3}$
satisfies
\begin{equation}
\label{d1}
 (e^{\rho}+e^{\tilde\rho})|B\big>_{3}=0,
\end{equation}
as in the flat case. 
The world-sheet variables transform
under the symmetry corresponding to $\int(G^{+}+\tilde G^{+})$ as 
\begin{eqnarray}
\label{1}
&&\delta_G\theta^\alpha=e^{\rho}d^\alpha,~~~~~~~~~
\delta_G\tilde\theta^\alpha=e^{\tilde\rho}\tilde
d^\alpha,\\\nonumber
&&\delta_G y^m=2ie^{\rho}d^\alpha\sigma^m_{\alpha\dot\alpha}\bar\theta^{\dot\alpha}
+2ie^{\tilde\rho}\tilde d^\alpha\sigma^m_{\alpha\dot\alpha}\!\tilde{~\bar\theta^{\dot\alpha}}
,\\\nonumber
&&\delta_G q^\alpha=-4\partial(e^{\rho}d^\alpha\bar{\theta}\bar{\theta}),
~~~~~~\delta_G \tilde q^\alpha=-4\tilde\partial
(e^{\tilde\rho}\tilde d^\alpha\,\tilde{\!\bar\theta}\,\tilde{\!\bar\theta}).
\end{eqnarray}
From (\ref{d3b}),(\ref{d1}) and  (\ref{1}), we determine the  equations for the
fermionic fields as
\begin{eqnarray}
\label{d4b}
\{F_{nm}\sigma^m_{\alpha\dot\alpha}
{\bar\sigma}^{n\dot\alpha\beta}(q^{\alpha}
-\tilde {q^\alpha})-{1\over \pi}(q^{\beta}+\tilde {q^\beta})\}|B\big>_3=0,\\\nonumber
\{F_{nm}\sigma^m_{\alpha\dot\alpha}{\bar\sigma}^{n\dot\alpha\beta}
(\theta^\alpha+{\tilde\theta^\alpha})
-{1\over \pi}(\theta^\beta-{\tilde\theta^\beta})\}|B\big>_3=0,\\\nonumber
(\bar d^{\dot\alpha}+\tilde {\bar d^{\dot\alpha}})|B\big>_3
=(\bar\theta^{\dot\alpha}-\tilde{\bar\theta^{\dot\alpha}})|B\big>_3=0.
\end{eqnarray}

\subsection{Boundary state for infinitely many D-instantons }
We consider the following boundary state as a generalization of eq.(\ref{b1}),
\begin{equation}
\label{b7}
 |B\big>=TrP\exp[-{1\over 4\pi}\int 
d\sigma\{(\partial+\tilde\partial)y^m\hat M_m
+(\partial\theta_\alpha+\tilde\partial{\tilde\theta_\alpha})\hat Q^\alpha
+(q^{\alpha}
+\tilde {q^\alpha})\hat\Theta_\alpha\}]|B\big>_{-1},
\end{equation} 
where $\hat M_m,\hat Q^\alpha,\hat\Theta_\alpha$ are
$\infty\times\infty$ matrices. $|B\big>_{-1}$ denotes  the boundary state for a D-instanton
satisfying
\begin{eqnarray}
\label{b3}
y^m|B\big>_{-1}=0,\\\nonumber
(q^\alpha-\tilde q^\alpha)|B\big>_{-1}=(\theta^\alpha+\tilde
\theta^\alpha)|B\big>_{-1}=0,\\\nonumber
(\bar d^{\dot\alpha}
+\tilde{\bar d^{\dot\alpha}})|B\big>_{-1}=(\bar\theta^{\dot\alpha}-\tilde{\bar
{\theta^{\dot\alpha}}})|B\big>_{-1}=0. 
\end{eqnarray}
As in the bosonic case, $\int d\sigma (\partial+\tilde\partial)y^m$, $\int d\sigma(q^{\alpha}
+\tilde {q^\alpha})$ and 
$\int d\sigma
(\partial\theta_\alpha+\tilde\partial{\tilde\theta_\alpha})$
should be identified as the vertex operators for a D-instanton.
In the hybrid formalism, an integrated vertex operator is given
by
\begin{equation}
\int dz G^-(G^+(V)).
\end{equation}
Here V is a function of $ \theta, \bar \theta$ for the vertex operator
in the    D-instanton case.
The  vertex operators 
$\int d\sigma (\partial+\tilde\partial)y^m$, $\int d\sigma(q^{\alpha}
+\tilde {q^\alpha})$,  
$\int d\sigma
(\partial\theta_\alpha+\tilde\partial{\tilde\theta_\alpha})$ 
can be obtained from $V=\theta^\alpha\sigma^m_{\alpha\dot\alpha}
\bar\theta^{\dot\alpha}$, $ \theta^{\alpha}\bar\theta\bar\theta$,  $\theta^{\alpha}$  respectively.
$\int d\sigma (\partial+\tilde\partial)y^m$ and $\int d\sigma(q^{\alpha}
+\tilde {q^\alpha})$ generate the translation of $y^m$ and 
$(\theta^\alpha+{\tilde\theta^\alpha})$ respectively and the vertex operator
$\int d\sigma  (\partial\theta_\alpha+\tilde\partial{\tilde\theta_\alpha})$ 
is a BRST exact operator generating the 
translation of a gauge degree of freedom. 
Therefore, $\hat M_m,\hat Q^\alpha,\hat\Theta_\alpha$ represent
a configuration of infinitely many D-instantons and (\ref{b7}) is the boundary 
state corresponding to such a configuration. 
Here we consider the  configuration in which $\hat M_m,\hat
Q^\alpha,\hat\Theta_\alpha$ satisfy
\begin{eqnarray} 
\label{c1}
[\hat M^n,\hat M^m]&=&i\theta^{nm},\\\nonumber
\{\hat\Theta_\alpha,\hat Q^\beta\}&=&{i\over4}\sigma^m_{\alpha\dot\alpha}
{\bar\sigma}^{n\dot\alpha\beta}\theta^{nm},
\end{eqnarray}
where $\theta^{nm}$ is self-dual.
The boundary state can be rewritten as the following path integral representation,
\begin{eqnarray}
\label{ub}
|B\big>=\int[dYd\chi d\vartheta]\prod_{\alpha,\gamma}
\chi_0^\gamma \{F_{mn}\sigma^n_{\alpha\dot\alpha}
{\bar\sigma}^{m\dot\alpha\beta}\vartheta_{\beta0}
+{1\over \pi}(\theta_{\alpha0}-{\tilde\theta_{\alpha0}})\}\hspace{5.2cm}\\\nonumber
\times\exp[{i\over2}\int d\sigma F_{mn}\{ Y^m\partial_{\sigma} Y^n
-{1\over2}\chi^\alpha\sigma^n_{\alpha\dot\alpha}
{\bar\sigma}^{m\dot\alpha\beta}\partial_{\sigma}\vartheta_\beta\}\hspace{4cm}
\\\nonumber
-{1\over 4\pi}\int 
d\sigma\{(\partial+\tilde\partial)y^mY_m
+(\partial\theta_\alpha+\tilde\partial{\tilde\theta_\alpha})\chi^\alpha
+(q^{\alpha}
+\tilde {q^\alpha})\vartheta_\alpha\}]|B\big>_{-1},
\end{eqnarray}
where $F_{mn}=(\theta^{nm})^{-1}$. 
$\vartheta,\chi$ are periodic with respect to $\sigma$ and 
 $\theta_{\alpha0},\vartheta_{\alpha0},\chi_{\alpha0}$ are the zero modes of 
$\theta_{\alpha}=\sum_{n}\theta_{\alpha n}e^{in\sigma},\vartheta_{\alpha}=\sum_{n}\vartheta_{\alpha n}e^{in\sigma},\chi_{\alpha}=\sum_{n}\chi_{\alpha n}e^{in\sigma}$ respectively. 
The eq.(\ref{ub}) is the generalization of eq.(\ref{12}) except for the
zero mode insertions of the fermionic fields.
In our case, we need such insertions to express the $Tr$ of the
Chan-Paton factors
in terms of the path integral representation.
Since the zero mode does not appear in the exponent of eq.(\ref{ub}),
we can understand the necessity of  $\chi_{\alpha0}$
 insertion. Although it is not clear why $\vartheta_{\alpha0}$ insertion
 is required, it becomes clear from the open string picture.
 
In this picture, the action in the background (\ref{c1}) can be guessed from 
(\ref{ub}) as 
\begin{eqnarray}
\label{a2}
 S=S_0+{i\over2}\int d\tau F_{mn}\{ Y^m\partial_{\tau} Y^n
-{1\over2}\chi^\alpha\sigma^n_{\alpha\dot\alpha}
{\bar\sigma}^{m\dot\alpha\beta}\partial_{\tau}\vartheta_\beta\}\hspace{4.5cm}\\\nonumber
+{i\over 4\pi}\int 
d\tau\{(\partial-\tilde\partial)y^mY_m
+(\partial\theta_\alpha-\tilde\partial{\tilde\theta_\alpha})\chi^\alpha
+(q^{\alpha}
-\tilde {q^\alpha})\vartheta_\alpha\},
\end{eqnarray}
where $S_0$ is the action of the free fields and $Y^m,\chi^\alpha,\vartheta_\beta$ 
are the Chan-Paton degrees of freedom on the boundaries.
Let us consider the case,  $F_{mn}\rightarrow\infty$. Redefining the fields appropriately,
the boundary action is given by
\begin{equation}
\label{a1}
\int d\tau  (Y^1\partial_{\tau} Y^2
+ Y^3\partial_{\tau} Y^4+
\chi^\alpha\partial_{\tau}\vartheta_\alpha).
\end{equation}
Canonically quantizing them, the Hilbert space is of the form
${\cal H_{\it Y}\otimes H_{\vartheta}}$. The basis of ${\cal H_{\it Y}}$
can be taken to be the eigenstates of $Y^1,Y^3$ and the basis of
${\cal H_{\vartheta}}$ can be taken to be $|0\big>,\vartheta_\alpha|0\big>,
\vartheta^\alpha\vartheta_\alpha|0\big>$ where $|0\big>$ is defined by $\chi_\alpha|0\big>=0$.
Since (\ref{a1})  corresponds to the background (\ref{c1})
with  $\theta^{mn}=0$ and represent infinitely many independent
D-instantons, we expect that the boundary state should be the one for
one D-instanton multiplied by $\infty$. 
By the following discussion, we can show that the insertion of $\vartheta_{\alpha}$ is
required in order to reproduce such result.
The simple trace over ${\cal H_{\vartheta}}$  is
\begin{equation}
 Tr_s {\cal O}={1\over2}\big<0|{\cal O}\vartheta^\alpha\vartheta_\alpha|0\big>
+\big<0|\vartheta^\alpha{\cal O}\vartheta_\alpha|0\big>
+{1\over2}\big<0|\vartheta^\alpha\vartheta_\alpha{\cal O}|0\big>.
\end{equation}
Since $\vartheta,\chi$
are anticommuting fields, this trace corresponds to the following path
integral,
\begin{equation}
 \int [d\vartheta d\chi]_{a.p}{\cal O}e^{\int d\tau\chi^\alpha\partial_{\tau}\vartheta_\alpha},
\end{equation}
where $\vartheta,\chi$ are antiperiodic with respect to $\tau$.
In order to obtain the periodic boundary condition, the trace should be  weighted
by $(-1)^{F}$
where $F$ is the fermion number, i.e.,
\begin{eqnarray}
Tr_s (-1)^{F}{\cal O}&=&
{1\over2}\big<0|(-1)^{F}{\cal O}\vartheta^\alpha\vartheta_\alpha|0\big>
+\big<0|\vartheta^\alpha(-1)^{F}{\cal O}\vartheta_\alpha|0\big>
+{1\over2}\big<0|\vartheta^\alpha\vartheta_\alpha(-1)^{F}{\cal O}|0\big>\hspace{1cm}\\\nonumber
&=&\int [d\vartheta d\chi]_{p}{\cal O}e^{\int
d\tau\chi^\alpha\partial_{\tau}\vartheta_\alpha}.
\end{eqnarray}
Here $\vartheta,\chi$ are now periodic with respect to $\tau$.
When ${\cal O}$  is the identity operator $1$, this trace is zero.
In order to make it non zero, we insert
$\chi^\alpha\chi_\alpha\vartheta^\beta\vartheta_\beta$
because
\begin{eqnarray}
  Tr_s \chi^\alpha\chi_\alpha\vartheta^\beta\vartheta_\beta(-1)^{F}1
&=&{1\over2}\big<0|\chi^\alpha\chi_\alpha\vartheta^\beta
\vartheta_\beta\vartheta^\gamma\vartheta_\gamma|0\big>
\\\nonumber
&&\hspace{1cm}-\big<0|\vartheta^\gamma\chi^\alpha\chi_\alpha\vartheta^\beta
\vartheta_\beta\vartheta_\gamma|0\big>+{1\over2}\big<0|\vartheta^\gamma\vartheta_\gamma
\chi^\alpha\chi_\alpha\vartheta^\beta\vartheta_\beta|0\big>\\\nonumber
&=&{1\over2}\big<0|\vartheta^\gamma\vartheta_\gamma
\chi^\alpha\chi_\alpha\vartheta^\beta\vartheta_\beta|0\big>\neq 0.
\end{eqnarray}
Since the trace over ${\cal H_{\it Y}}$ can be found to be infinite, we can obtain
the desire result.
In the closed string picture, this trace leads to the following boundary state,
\begin{eqnarray}
\int[dYd\chi d\vartheta]\prod_{\alpha,\gamma}
\chi_0^\gamma\vartheta^\alpha_0 e^{\int d\sigma  (Y^1\partial_{\sigma} Y^2
+ Y^3\partial_{\sigma} Y^4+
\chi^\alpha\partial_{\sigma}\vartheta_\alpha})
|B\big>_{-1},
\end{eqnarray}
which  coincides with the
boundary state for infinitely many independent D-instantons.
For the general background where the action is given by (\ref{a2}),
the insertion is modified as $\chi p_\chi$ 
where $p_\chi$ is the canonical conjugate of
$\chi$, i.e. 
$p_{\chi^\alpha}={1\over \pi}
(\theta_\alpha-{\tilde\theta_\alpha})+F_{mn}\sigma^n_{\alpha\dot\alpha}
{\bar\sigma}^{m\dot\alpha\beta}\vartheta_\beta$.

We would like  to show that this boundary state corresponds to the boundary state for
a D$3$-brane in a constant self-dual gauge field $F_{mn}$ 
background. Indeed, $|B\big>$ satisfies the following equations for the non
zero mode,
\begin{eqnarray}
\label{e3}
0&=&\int[dYd\chi d\vartheta]{\delta\over{\delta Y^l}}
\exp[{i\over 2}\int d\sigma F_{mn}\{ Y^m\partial_{\sigma} Y^n
-{1\over 2}\chi^\alpha\sigma^n_{\alpha\dot\alpha}
{\bar\sigma}^{m\dot\alpha\beta}\partial_{\sigma}\vartheta_\beta\}\hspace{3cm}
\\\nonumber
&&\hspace{2cm}-{1\over 4\pi}\int 
d\sigma\{(\partial+\tilde\partial)y^mY_m
+(\partial\theta^\alpha+\tilde\partial{\tilde\theta^\alpha})\chi_\alpha
+(q^{\alpha}+
\tilde {q^\alpha})\vartheta_\alpha\}]|B\big>_{-1}\\\nonumber
&=&
\{{1\over 4\pi}(\partial+\tilde\partial)y^l+F_{ln} (\partial-\tilde\partial)y^n\}|B\big>,
\\\nonumber
0&=&\int[dYd\chi d\vartheta]{\delta\over{\delta \vartheta_\gamma}}
\exp[{i\over2}\int d\sigma F_{mn}\{ Y^m\partial_{\sigma} Y^n
-{1\over 2}\chi^\alpha\sigma^n_{\alpha\dot\alpha}
{\bar\sigma}^{m\dot\alpha\beta}\partial_{\sigma}\vartheta_\beta\}\hspace{3cm}
\\\nonumber
&&\hspace{2cm}-{1\over 4\pi}\int 
d\sigma\{(\partial+\tilde\partial)y^mY_m
+(\partial\theta^\alpha+\tilde\partial{\tilde\theta^\alpha})\chi_\alpha
+(q^{\alpha}+
\tilde {q^\alpha})\vartheta_\alpha\}]|B\big>_{-1}\\\nonumber
&=&\{F_{mn}\sigma^n_{\alpha\dot\alpha}
{\bar\sigma}^{m\dot\alpha\gamma}(q^{\alpha}
-\tilde {q^\alpha})-{1\over \pi}(q^\gamma+\tilde {q^\gamma})\}|B\big>,\\\nonumber
0&=&\int[dYd\chi d\vartheta]{\delta\over{\delta \chi^\gamma}}
\exp[{i\over2}\int d\sigma F_{mn}\{ Y^m\partial_{\sigma} Y^n
-{1\over2}\chi^\alpha\sigma^n_{\alpha\dot\alpha}
{\bar\sigma}^{m\dot\alpha\beta}\partial_{\sigma}\vartheta_\beta\}\hspace{3cm}
\\\nonumber
&&\hspace{2cm}-{1\over 4\pi}\int 
d\sigma\{(\partial+\tilde\partial)y^mY_m
+(\partial\theta^\alpha+\tilde\partial{\tilde\theta^\alpha})\chi_\alpha
+(q^{\alpha}+
\tilde {q^\alpha})\vartheta_\alpha\}]|B\big>_{-1}\\\nonumber
&=& 
\{F_{mn}\sigma^n_{\gamma\dot\alpha}{\bar\sigma}^{m\dot\alpha\alpha}
(\partial\theta_\alpha-\tilde\partial{\tilde\theta_\alpha})
+{1\over \pi}(\partial\theta_\gamma+\tilde\partial{\tilde\theta_\gamma})\}|B\big>.
\end{eqnarray}
Thus, we have shown $|B\big>$ satisfies almost all in the equations
(\ref{d4b}) except for the ones involving $\theta_0,\tilde\theta_0$.
Since (\ref{ub}) includes the following fact,
\begin{eqnarray}
&&\int[d\vartheta_0]\prod_{\alpha}\{
F_{mn}\sigma^n_{\alpha\dot\alpha}
{\bar\sigma}^{m\dot\alpha\beta}\vartheta_{\beta0}
+{1\over \pi}
(\theta_{\alpha0}-{\tilde\theta_{\alpha0}})\}e^{(q^\gamma_0+\tilde q^\gamma_0)\vartheta_{\gamma0}}
|B\big>_{-1}\\\nonumber
&=&\int[d\vartheta_0]\prod_{\alpha}\{
F_{mn}\sigma^n_{\alpha\dot\alpha}
{\bar\sigma}^{m\dot\alpha\beta}(\theta_{\beta0}+{\tilde\theta_{\beta0}})
+{1\over \pi}
(\theta_{\alpha0}-{\tilde\theta_{\alpha0}})\}e^{(q^\gamma_0+\tilde q^\gamma_0)\vartheta_{\gamma0}}
|B\big>_{-1}
\\\nonumber
&=&\prod_{\alpha,\gamma}\{
F_{mn}\sigma^n_{\alpha\dot\alpha}
{\bar\sigma}^{m\dot\alpha\beta}(\theta_{\beta0}+{\tilde\theta_{\beta0}})
+{1\over \pi}(\theta_{\alpha0}-{\tilde\theta_{\alpha0}})\}(q^\gamma_0+\tilde q^\gamma_0)
|B\big>_{-1},
\end{eqnarray}
we can show that $|B\big>$ satisfies the equations
(\ref{d4b}).
Thus, the boundary state $|B\big>$ coincides with for
a D$3$-brane in a constant self-dual gauge field $F_{mn}$ 
background.

\section{Conclusions and discussions}
 In this paper,
we express the boundary state 
for a D$3$-brane in a constant self-dual gauge 
field background in terms of the  boundary state for 
infinitely many D-instantons in  the hybrid formalism.
Therefore, a D$3$-brane in a constant self-dual gauge 
field background can be regarded as a configuration of 
infinitely many D-instantons in  the hybrid formalism.
In order to discuss the above statement more exactly, we need to study
the open string theory  corresponding to  infinitely many D-instantons in the
background (\ref{c1}). As mentioned above, we should pay  attention to
the zero mode insertions. 
 
In the bosonic string, we can express the boundary state for a D$p$-brane
in a constant gauge field background
in terms of  the boundary state for D$(p-2k)$-branes.
In the hybrid formalism,  we have discussed the case where
$p=3, k=2$ and the constant gauge field is self-dual.
Let us consider other cases where
$p-2k=-1$ and the constant gauge field is general. In such cases, 
the equations for the field  
$(\bar d^{\dot\alpha}+\tilde {\bar d^{\dot\alpha}})$ satisfied by 
the boundary state for  a D$p$-brane
are different from those for  D-instantons.
In order to modify such boundary conditions, we should add 
$(\bar q^{\dot\alpha}-\tilde {\bar q^{\dot\alpha}})$ in (\ref{b7}).
These operators are not  free fields and do not commute with $y^m$ and so on. 
Thus we can not obtain the equations as   (\ref{e3}) easily.
For this reason, it is difficult to decide the commutation relations 
of matrices as (\ref{c1}) to express the boundary state for a  D$p$-brane
in terms of the boundary state for  D-instantons. For a D$3$-brane in a constant
 self-dual gauge field background, since the equations 
for the field 
$(\bar d^{\dot\alpha}+\tilde {\bar d^{\dot\alpha}})$ are same as
those for D-instantons, we can express the boundary state for  a D$3$-brane
in terms of the boundary state for  D-instantons.

Our construction is the first step to understand the general non(anti)commutativity by lower
dimensional D-branes.
  We would like to consider a D-brane in a  graviphoton
background using our construction in a separate publication.

\section*{ Acknowledgments}
We would like to thank N. Ishibashi for useful discussions and careful reading of
the manuscript.


\begin{thebibliography}{10}
 
\bibitem{aas}
N.~Ardalan, H.~Arfaei and M.~M.~Sheikh-Jabbari,
\JL{J.\ High\ Energy\ Phys.,9902,1999,016};
hep-th/9810072

\bibitem{ch}
C.-S.~Chu and P.-M.~Ho,
\NP{B550,1999,151};
hep-th/9812219

\bibitem{is1}
N.~Ishibashi,
\NP{B539,1999,107};
hep-th/9804163

\bibitem{is2}
N.~Ishibashi,
hep-th/9909176

\bibitem{sw}
N.~Seiberg and E.~Witten,
\JL{J.\ High\ Energy\ Phys.,9909,1999,032};
hep-th/9908142

\bibitem{ok}
K.~Okuyama,
\JL{J.\ High\ Energy\ Phys.,0003,2000,016};
hep-th/9910138

\bibitem{si}
N.~Seiberg,
\JL{J.\ High\ Energy\ Phys.,0306,2003,010}; 
hep-th/0305248

\bibitem{besi}
N.~Berkovits and N.~Seiberg, 
\JL{J.\ High\ Energy\ Phys.,0307,2003,010};
hep-th/0306226

\bibitem{bgn}
J.~de Boer, P.~A.~Grassi and P.~van Nieuwenhuizen,
\PL{B574,2003,98};
hep-th/0302078

\bibitem{ov}
H.~Ooguri and C.~Vafa, 
\JL{Adv.\ Theor.\ Math.\ Phys.,7,2003,053};
hep-th/0302109

\bibitem{be}
N.~Berkovits,
hep-th/9604123

\bibitem{mi}
D.~Mikulovi\' c,
\JL{J.\ High\ Energy\ Phys.,0405,2004,077};
hep-th/0403290

\bibitem{pe}
J.~-H.~Park, 
\JL{J.\ High\ Energy\ Phys.,0309,2003,046};
hep-th/0307060

\bibitem{kkm}
H.~Kawai, T.~Kuroki and T.~Morita,
\NP{B664,2003,185};
hep-th/0303210

\bibitem{hiu}
M.~Hatsuda, S.~Iso and H.~Umetsu,
\NP{B671,2003,217};
hep-th/0306251

\bibitem{st}
Y.~Shibusa and T.~Tada,
\PL{B579,2004,211};
hep-th/0307236



\end{thebibliography}
\end{document}